\documentclass[preprint,12pt]{elsarticle}
\usepackage{amsmath,amssymb,amsfonts}
\usepackage{graphicx}
\usepackage{epsfig}
\usepackage{slashed}
\usepackage{a4wide}
\usepackage{fancyhdr}
\usepackage{subfigure}

\allowdisplaybreaks[1]

\begin{document}

\newcommand{\derivl}{\stackrel{\leftarrow}{\partial}}
\newcommand{\derivr}{\stackrel{\rightarrow}{\partial}}

\newcommand{\evl}{e^{\frac{i\stackrel{\leftarrow}{\partial}_{\alpha}v^{\alpha}+m}{\Lambda}}}
\newcommand{\evr}{e^{\frac{-v^{\alpha}i\stackrel{\rightarrow}{\partial}_{\alpha}+m}{\Lambda}}}
\newcommand{\esl}{e^{\frac{i\stackrel{\leftarrow}{\partial}_{\alpha}v^{\alpha}+m}{\Lambda}}}
\newcommand{\esr}{e^{\frac{-v^{\alpha}i\stackrel{\rightarrow}{\partial}_{\alpha}+m}{\Lambda}}}

\newcommand{\err}{e^{\frac{-v^{\alpha}i\stackrel{\rightarrow}{\partial}_{\alpha}+m}{\Lambda}}}
\newcommand{\Gv}{\frac{g_{\omega}}{\Lambda}}
\newcommand{\Gr}{\frac{g_{\rho}}{\Lambda}}
\newcommand{\Gs}{\frac{g_{\sigma}}{\Lambda}}

\newcommand{\evnml}{e^{\frac{E}{\Lambda}}}
\newcommand{\esnml}{e^{\frac{E}{\Lambda}}}

\newcommand{\evnm}{e^{-\frac{E-m}{\Lambda}}}
\newcommand{\esnm}{e^{-\frac{E+m}{\Lambda}}}

\newcommand{\Gva}{\frac{g_{\omega}}{\Lambda}}
\newcommand{\Gsa}{\frac{g_{\sigma}}{\Lambda}}

\newcommand{\Gvb}{\frac{g_{\omega}}{ (\Lambda)^{2}}}
\newcommand{\Gsb}{\frac{g_{\sigma}}{ (\Lambda)^{2}}}

\newcommand{\NPA}{Nucl.~Phys.~}
\newcommand{\PR}{Phys.~Rep.~}
\newcommand{\PL}{Phys.~Lett.~}
\newcommand{\PRC}{Phys.~Rev.~}
\newcommand{\PRL}{Phys.~Rev.~Lett.~}
\newcommand{\EPJ}{Eur.~Phys.~J.~}
\begin{frontmatter}

\title{Energy Dependent Isospin Asymmetry in Mean-Field Dynamics}

\author{T.~Gaitanos, M.~Kaskulov}
\address{ Institut f\"ur Theoretische Physik, Universit\"at Giessen,
             D-35392 Giessen, Germany}

\begin{abstract}
The Lagrangian density of relativistic mean-field (RMF) theory 
with non-linear derivative (NLD) interactions is applied to 
isospin asymmetric nuclear matter. We study the symmetry energy and 
the density and energy dependences of nucleon selfenergies.  
At high baryon densities a soft symmetry energy  is obtained. 
The energy dependence of the isovector selfenergy 
suppresses the Lane-type  optical potential with 
increasing energy and predicts a
$\rho$-meson induced mass splitting 
between protons and neutrons in isospin asymmetric matter. 
\end{abstract}

\end{frontmatter}

\date{\today}

\section{\label{sec1}Introduction}

The study of isospin asymmetric nuclear matter is closely related to a few issues 
which are presently the subjects of ongoing theoretical and experimental researches. 
At first, the isospin dependence of the nuclear equation of state (EoS) shows up
in the density dependence of the symmetry energy. A quantity which drives 
the structure of neutron rich nuclei and at supra-high densities the physics of 
supernovae explosions and the structure of compact neutron stars~\cite{ditoro,baoan}. 
For instance, the stiffness~\cite{stiff}
of the symmetry energy is directly related to the 
proton-fraction in $\beta$-equilibrated neutron stars, which is limited by the 
onset of the URCA-process~\cite{ns1}. However, the symmetry energy is empirically 
known up to saturation densities, 
where it can be determined in nuclear structure studies from Giant 
and Pigmy resonances and from neutron skin 
thickness~\cite{a4emp}. 
In the laboratory the  symmetry energy 
can be explored in heavy ion collisions, where the knowledge of the high density
EoS is mandatory~\cite{fopi1,fopi2} for the description of the 
collision dynamics~\cite{dani2,aichelin,gibuu,Cassing:1999es,bertsch,greiner} 
within transport theory~\cite{dani1,botermans}. Different heavy ion observables, {\it e.g.},  
collective transverse (in-plane) and elliptic (out-of-plane) flows~\citep{andronic} 
of particles and/or light fragments, as well as subthreshold strangeness 
production~\cite{fuchshic} are sensitive to the stiffness of the EoS. 
Furthermore, the isospin 
tracing~\cite{isotrac}, isoscaling~\cite{isoscal}, 
collective isospin flows~\cite{isoflows1,isoflows2,isoflows3,isoflows4} and 
isospin ratios of produced mesons~\cite{mesrat0,mesrat1,mesrat2,mesrat3} 
represent plausible tools to constrain the isovector EoS 
at high densities and energies. 
Moreover, heavy ion collisions allow to probe not only the density of created matter, 
but also the energy dependence of the nuclear potential~\cite{cassing,danimom,sahu}.

Theoretically, the density and energy dependence of the isovector EoS 
beyond the saturation density is still controversial. In non-relativistic 
Skyrme models both possibilities of a soft and a 
stiff symmetry energies~\cite{ditoro,baoan,isoflows2,isoflows3} can be obtained. 
In the RMF  models 
the symmetry energy is determined by the Lorentz-structure 
of the isovector potential~\cite{ditoro}, which is described 
by the vector-isovector $\rho$-field, 
or by $\rho$- and an attractive scalar $\delta$-fields~\cite{kubis,liu}. 
In both cases, the symmetry energy rises with increasing baryon 
density leading to a stiff isovector EoS at high densities. 
The models with density dependent (DD) couplings~\cite{lenske,typel1,ddOthers,dde}
predict a soft symmetry energy~\cite{typel2}, while the Dirac-Brueckner-Hartree-Fock 
(DBHF) calculations~\cite{dbfuchs,dbfuchs_new} and other RMF models 
with complementary exchange (Fock) terms (density-dependent relativistic 
Hartree-Fock (RHF)~\cite{rhf1,rhf2}) favour again a stiff one.

Concerning the energy dependence, the Schr\"odinger equivalent optical potential 
in symmetric matter is well known phenomenologically~\cite{dp}. However, different 
empirical analyses give different predictions for the energy dependence of 
isovector Lane-potential~\cite{isodp1,isodp1b,isodp1c,isodp2,isodp2b}. The microscopic DBHF~\cite{dbfuchs} and 
BHF~\cite{bhf1,bhf2} models predict a rather constant optical potential at 
low momenta with a strong decrease at high energies. In Skyrme approaches both cases 
of a strong increase and a strong decrease of the Lane-potential as a function 
of energy can be realized~\cite{baoan}. The DD models do not contain any explicit 
energy dependence, except if one modifies the original DD formalism~\cite{lenske,typel1} 
with additional interaction terms~\cite{typel2}, however, with the cost of additional 
parameters. Note that, the energy dependence of the optical potential has been so 
far a crucial problem in standard RMF, where the energy dependence appears only due 
to relativistic effects. 

Another question concerns the quasi-particle properties of nucleons in the nuclear 
matter. The isospin-dependent forces may result in a splitting between the in-medium 
masses of protons and neutrons which can strongly influence the dynamics in 
heavy-ion collisions~\cite{isoflows2,isoflows3}. 
However, it is still under debate, whether 
$m^{*}_{n} < m^{*}_{p}$ or $m^{*}_{n} > m^{*}_{p}$ holds in neutron rich matter. 
In non-relativistic approaches the in-medium particle mass is described through 
the energy dependence of the real part of the phenomenological mean-field~\cite{jami}. 
Again, the Skyrme-like interactions predict both possibilities~\cite{baoan} 
for the non-relativistic in-medium mass~\cite{jami}. 
In relativistic models the in-medium dependence of the particle mass arises from 
the Lorentz-scalar character of the relativistic mean-field, i.e., from the scalar 
selfenergy~\cite{jami}. It is often denoted as "effective mass" or "Dirac mass", and these 
definitions we will use in this work. 
As in the non-relativistic case, same situation holds for the DBHF models where
some calculations  predict for the Dirac mass 
$m^{*}_{n} < m^{*}_{p}$~\cite{dbfuchs} and others 
$m^{*}_{n} > m^{*}_{p}$~\cite{dbothers,dbothersb}. 
In RMF models the mass splitting  
is merely fixed by the Lorentz structure of the isovector mean-field. 
Due to the Lorentz-vector character of the $\rho NN$-interaction the minimal 
$\sigma\omega\rho$-models do not 
generate any in-medium mass splitting, except if one in addition introduces the 
Lorentz-scalar $\delta$-field in the isovector channel~\cite{kubis,liu}, 
or if one includes explicitly 
exchange (Fock) terms in the relativistic formalism~\cite{rhf1,rhf2}. 
Including the $\delta$-field one gets $m^{*}_{n} < m^{*}_{p}$, 
while in the RHF models both cases $m^{*}_{n} \gtrless m^{*}_{p}$ are possible. 
So far these were the only possibilities to generate the mass splitting in the 
relativistic mean field theory~\cite{kubis,liu,typel2,rhf1,rhf2}. 

In this work we study the isovector interactions in asymmetric nuclear matter.
Our studies are based on the non-linear derivative (NLD) model  
to RMF proposed in~\cite{nld}. This approach describes remarkably well the bulk 
properties of nuclear matter and, 
in particular, the density dependence of the nuclear 
EoS and the energy dependence of the proton-nucleus~\cite{nld}
as well as the antiproton-nucleus Schr\"odinger equivalent 
optical potentials~\cite{antinld}.  The present paper is organized as
follows. In Section~2 we briefly outline the NLD model. In
Section~3 the approach is applied to isospin symmetric and isospin asymmetric
nuclear matter. The numerical realization and details of the calculations 
are described in Section~4. The results and discussions are presented in Section~5. We
finally end up with the summary in Section~6.

\section{\label{sec2}Non-Linear Derivative (NLD) Model  }


The NLD approach is based on the Lagrangian density of relativistic hadrodynamics 
(RHD)~\cite{rhd1,rhd2} and describes the in-medium interaction of nucleons 
through the  exchange of auxiliary scalar-isoscalar $\sigma$,  
vector-isoscalar $\omega^{\mu}$ and vector-isovector $\vec{\rho\,}^{\mu}$
meson fields. The Lagrangian density is given by
\begin{eqnarray}
{\cal L}& = & \frac{1}{2}
\left[
	\overline{\Psi}i\gamma_{\mu}\partial^{\mu}\Psi
	- 
	(i\partial^{\mu}\overline{\Psi}) \gamma_{\mu}\Psi
\right]
- \overline{\Psi}\Psi m
\nonumber\\
& - & 
\frac{1}{2}m^{2}_{\sigma}\sigma^{2}
+\frac{1}{2}\partial_{\mu}\sigma\partial^{\mu}\sigma
+\frac{1}{2}m^{2}_{\omega}\omega_{\mu} \omega^{\mu} 
-\frac{1}{4}F_{\mu\nu}F^{\mu\nu}
+ \frac{1}{2}m^{2}_{\rho}\vec{\rho\,}_{\mu}\vec{\rho\,}^{\mu} 
-\frac{1}{4}\vec{G\,}_{\mu\nu}\vec{G\,}^{\mu\nu}
\nonumber\\ 
& + &
{\cal L}_{int}
\quad ,
\label{NDC-free}
\end{eqnarray}
where the first two terms describe the (symmetrized) Lagrangian 
for the nucleon field $\Psi=(\Psi_{p},\Psi_{n})^{T}$ with bare mass $m$ and 
the second line contains the standard Lagrangian densities for 
$\sigma$, $\omega^{\mu}$ and $\vec{\rho\,}^{\mu}$ fields with masses 
$m_\sigma$, $m_\omega$ and $m_\rho$, respectively. The 
strength tensors are defined as 
$F^{\mu\nu}=\partial^{\mu}\omega^{\nu}-\partial^{\nu}\omega^{\mu}$ 
and 
$\vec{G\,}^{\mu\nu}=\partial^{\mu}\vec{\rho\,}^{\nu}-\partial^{\nu}\vec{\rho\,}^{\mu}$ 
for the vector-isoscalar  and vector-isovector fields, correspondingly.  

In Eq.~(\ref{NDC-free}) the last term ${\cal L}_{int}$ contains the interaction between 
the meson and nucleon fields. 
In the conventional RHD~\citep{rhd1,rhd2} the meson fields couple to 
the nucleons  via the corresponding Lorentz structures
$\overline{\Psi}\Psi\sigma$, 
-$\overline{\Psi}\gamma^{\mu}\Psi\omega_{\mu}$ and 
-$\overline{\Psi}\gamma^{\mu}\vec{\tau}\Psi\vec{\rho}_{\mu}$. 
Such minimal Walecka-type interactions describe successfully the static 
properties of nuclear matter around saturation density. However, they are not 
sufficient for the description of dynamical situations which occur in 
nucleon-nucleus and heavy-ion collisions, where this minimal model results in 
a wrong energy dependence of the mean-field. This has been a crucial problem 
of the RMF models which already attracted much attention in the 
past~\cite{sahu,typel2,Maruyama}. 
A possible solution to this problem has been suggested in~\cite{cassing}, where 
momentum-dependent form factors which suppress the high momentum components of
fields were introduced. This idea has been followed in the NLD 
model~\cite{nld} where it was generalized in a manifestly covariant way. 

The NLD interaction Lagrangian introduces the non-linear derivative terms into the 
conventional RHD meson-nucleon vertices. In particular, the Lagrangian density depends 
on higher-order partial derivatives of the nucleon field $\Psi$
\begin{equation}
\label{Lf}
{\cal L}_{int} \equiv 
{\cal L}(\Psi, \, \partial_{\alpha}\Psi, 
\,\, \partial_{\alpha}\partial_{\beta}\Psi, 
\,\, \cdots, 
\,\, \overline{\Psi}, 
\,\, \partial_{\alpha}\overline{\Psi}, 
\,\, \partial_{\alpha}\partial_{\beta}\overline{\Psi}, 
\,\, \cdots)
\end{equation}
with the symmetrized meson-nucleon interactions given by ($\tau$ denotes the
isospin operator)
\begin{align}
{\cal L}_{int}  & = 
\frac{g_{\sigma}}{2}
	\left[
	\overline{\Psi}
	\, \stackrel{\leftarrow}{{\cal D}}
	\Psi\sigma
	+\sigma\overline{\Psi}
	\, \stackrel{\rightarrow}{{\cal D}}
	\Psi
	\right]
\nonumber\\
& 
-  \frac{g_{\omega}}{2}
	\left[
	\overline{\Psi}
	 \, \stackrel{\leftarrow}{{\cal D}}
	\gamma^{\mu}\Psi\omega_{\mu}
	+\omega_{\mu}\overline{\Psi}\gamma^{\mu}
	\, \stackrel{\rightarrow}{{\cal D}}
	\Psi
	\right]
\nonumber\\
& 
-  \frac{g_{\rho}}{2}
	\left[
	\overline{\Psi}
	 \, \stackrel{\leftarrow}{{\cal D}}
	\gamma^{\mu}\vec{\tau}\Psi\vec{\rho}_{\mu}
	+\vec{\rho}_{\mu}\overline{\Psi}\vec{\tau}\gamma^{\mu}
	\, \stackrel{\rightarrow}{{\cal D}}
	\Psi
	\right]
\quad ,
\label{NDC}
\end{align}
with obvious notations for the fields and their couplings. The new operator ${\cal D}$ acts 
on the nucleon fields and is the generic non-linear function of partial derivatives. 
In the model of Ref.~\cite{nld} a simple exponential cut-off function has been assumed
\begin{equation}
\label{ope}
\stackrel{\rightarrow}{{\cal D}}
:= \exp{\left(\frac{-v^{\alpha}i\stackrel{\rightarrow}{\partial}_{\alpha}+m}{\Lambda}\right)} =  \sum_{n=0}^{\infty}~
\frac{(-v^{\alpha}\;i\!\stackrel{\rightarrow}{\partial}_{\alpha}/\Lambda)^{n}}{n!} 
e^{m/\Lambda}
\,,
\end{equation}
\begin{equation}
\stackrel{\leftarrow}{{\cal D}}
:= \exp{\left(\frac{i\stackrel{\leftarrow}{\partial}_{\alpha}v^{\alpha}+m}{\Lambda}\right)} =  \sum_{n=0}^{\infty}~
\frac{(i\!\stackrel{\leftarrow}{\partial}_{\alpha}v^{\alpha}/\Lambda)^{n}}{n!} 
e^{m/\Lambda}
\,. \label{ope2}
\end{equation}
In Eqs.~(\ref{ope}) and~(\ref{ope2}) 
$v^{\alpha}$ denotes an auxiliary dimensionless  $4$-vector and $\Lambda$ stands for the 
cut-off parameter. A way to realize the meaning of the cut-off function 
is the presence of the momentum (or energy) dependent form factors in the 
one-boson-exchange models, which cut the high energy behaviour of the meson-exchange vertices~\cite{obe,dbhf1}. Furthermore, 
in the limiting case of $\Lambda\rightarrow\infty$ the standard RHD (or Walecka) model is retained. 


The Euler-Lagrange equations and the Noether currents 
are derived from variational principles which are applied to the
NLD Lagrangian containing higher order derivatives 
of the spinor fields $\Psi$ and $\overline{\Psi}$. The generalized 
Euler-Lagrange equations read~\cite{nld} ($\phi=\Psi,\overline{\Psi})$
\begin{align}\label{Euler}
\frac{\partial{\cal L}}{\partial\phi}
-
 \partial_{\alpha_{1}}\frac{\partial{\cal L}}{\partial(\partial_{\alpha_{1}}\phi)}
+
 \partial_{\alpha_{1}}\partial_{\alpha_{2}}\frac{\partial{\cal L}}{\partial(\partial_{\alpha_{1}}\partial_{\alpha_{2}}\phi)}
& + \cdots  + \\
& (-)^{n}\partial_{\alpha_{1}}\partial_{\alpha_{2}}\cdots\partial_{\alpha_{n}}
\frac{\partial{\cal L}}
{\partial(\partial_{\alpha_{1}}\partial_{\alpha_{2}}\cdots\partial_{\alpha_{n}}\phi)}=
0 \nonumber
\quad .
\end{align}

The invariance of the NLD Lagrangian under global phase  transformations results 
in the following Noether current $J^{\mu}$~\cite{nld}
\begin{align}
J^{\mu} = & 
\left[ 
 \frac{\partial{\cal L}}{\partial(\partial_{\mu}\phi)} 
 - 
 \partial_{\beta}
 \frac{\partial{\cal L}}{\partial(\partial_{\mu}\partial_{\beta}\phi)}
 +
 \partial_{\beta}\partial_{\gamma}
 \frac{\partial{\cal L}}{\partial(\partial_{\mu}\partial_{\beta}\partial_{\gamma}\phi)}
 \mp \cdots 
 \hspace*{3.6cm}
 \right]\phi   
\nonumber\\
+ & \left[ 
 \hspace*{2.25cm}
 \frac{\partial{\cal L}}{\partial(\partial_{\mu}\partial_{\beta}\phi)}
 - 
 \hspace*{0.4cm}
 \partial_{\gamma}
 \frac{\partial{\cal L}}{\partial(\partial_{\mu}\partial_{\beta}\partial_{\gamma}\phi)}
 \pm \cdots 
 \hspace*{3.6cm}
 \right]\partial_{\beta}\phi
\nonumber\\
+ & \left[ 
 \hspace*{5.25cm}
 \frac{\partial{\cal L}}{\partial(\partial_{\mu}\partial_{\beta}\partial_{\gamma}\phi)}
- 
 \partial_{\delta}
 \frac{\partial{\cal L}}{\partial(\partial_{\mu}\partial_{\beta}\partial_{\gamma}\partial_{\delta}\phi)}
 \pm \cdots 
 \hspace*{0.25cm}
 \right]\partial_{\beta}\partial_{\gamma}\phi 
\nonumber\\
+ & \cdots 
\quad .
\label{Noether-Current}
\end{align}
Furthermore, the invariance under space-time translations gives the energy-momentum 
tensor $T^{\mu\nu}$~\cite{nld}
\begin{align}
T^{\mu\nu} =&
 \left[
   \frac{\partial{\cal L}}{\partial(\partial_{\mu}\phi)}
 - \partial_{\beta}
   \frac{\partial{\cal L}}{\partial(\partial_{\mu}\partial_{\beta}\phi)}
 + \partial_{\beta}\partial_{\gamma}
   \frac{\partial{\cal L}}{\partial(\partial_{\mu}\partial_{\beta}\partial_{\gamma}\phi)}
 \mp \cdots
 \hspace*{3.4cm}
 \right]\partial^{\nu}\phi
\nonumber\\
+ & \left[
   \hspace*{2.25cm}
   \frac{\partial{\cal L}}{\partial(\partial_{\mu}\partial_{\beta}\phi)}
 - \hspace*{0.4cm}
 \partial_{\gamma}
   \frac{\partial{\cal L}}{\partial(\partial_{\mu}\partial_{\beta}\partial_{\gamma}\phi)}
 \pm \cdots
 \hspace*{3.4cm}
 \right]\partial_{\beta}\partial^{\nu}\phi
\nonumber\\
+ & \left[
   \hspace*{5.25cm}
   \frac{\partial{\cal L}}{\partial(\partial_{\mu}\partial_{\beta}\partial_{\xi}\phi)}
 - \partial_{\gamma}
   \frac{\partial{\cal L}}{\partial(\partial_{\mu}\partial_{\gamma}\partial_{\beta}\partial_{\xi}\phi)}
 \pm \cdots
 \right]\partial_{\beta}\partial_{\xi}\partial^{\nu}\phi 
\nonumber\\  
+ & \cdots 
\nonumber\\ 
- & g^{\mu\nu}{\cal L} 
\quad .
\label{Noether}
\end{align}
The Euler-Lagrange equations~(\ref{Euler}) 
and the Noether theorems~(\ref{Noether-Current}) and~(\ref{Noether}) contain now an 
infinite series of higher order derivatives of the nucleon field. 

It was shown in~\cite{nld} that in mean-field approximation (up to derivatives of the 
meson fields which in any case vanish in nuclear matter) these infinite series can be 
resumed exactly. Indeed, the application of the generalized Euler-Lagrange 
equations~(\ref{Euler}) to the full Lagrangian density (\ref{NDC}) with respect to the 
field $\overline{\Psi}$ leads to the following Dirac equation~\cite{nld}
\begin{equation}
\left[
	\gamma_{\mu}(i\partial^{\mu}-\Sigma^{\mu}) - 
	(m-\Sigma_{s})
\right]\Psi = 0
\; ,
\label{Dirac}
\end{equation}
with the selfenergies $\Sigma^{\mu}$ and $\Sigma_{s}$ given by 
\begin{eqnarray}
\Sigma^{\mu} & = & g_{\omega}\omega^{\mu}\evr + g_{\rho}\vec{\tau} \cdot \vec{\rho\,}^{\mu}\err \label{Sigma_s}\,,\\
\Sigma_{s} & = & g_{\sigma}\sigma\esr
\;. \label{Sigma}
\end{eqnarray}
Both Lorentz-components of the selfenergy, $\Sigma^{\mu}$ and $\Sigma_{s}$, 
show a linear dependence with respect to the meson fields 
$\vec{\rho\,}^{\mu}$, $\omega^{\mu}$ and $\sigma$, as in the standard RMF. 
However, they contain an additional non-linear dependence on the partial derivatives. 

The following Proca and Klein-Gordon equations for the meson fields are obtained
\begin{eqnarray}
\partial_{\mu}F^{\mu\nu} + m_{\omega}^{2}\omega^{\nu} &=& 
\frac{1}{2}g_{\omega}
\left[
	\overline{\Psi}\evl \gamma^{\nu}\Psi + \overline{\Psi}\gamma^{\nu}\evr \Psi
\right],
\label{omega_meson} \\
\partial_{\mu}\vec{G}^{\mu\nu} + m_{\rho}^{2}\vec{\rho\,}^{\nu} &=& 
\frac{1}{2}g_{\rho}
\left[
	\overline{\Psi}\evl \gamma^{\nu}\vec{\tau}\Psi + \overline{\Psi}\vec{\tau}\gamma^{\nu}\evr \Psi
\right],
\label{rho_meson} \\
\partial_{\mu}\partial^{\mu}\sigma + m_{\sigma}^{2}\sigma &=& 
\frac{1}{2}g_{\sigma}
\left[
	\overline{\Psi}\evl \Psi + \overline{\Psi}\evr \Psi
\right]
\;, \label{sigma_meson}
\end{eqnarray}
where Eqs.~(\ref{omega_meson},~\ref{rho_meson},~\ref{sigma_meson}) show a similar 
form as in conventional RMF ($\Lambda\to\infty$), except of the highly non-linear 
structure in the source terms. 

\section{\label{sec3}Isospin-asymmetric nuclear matter}

We apply now the NLD model to isospin-asymmetric nuclear matter. 
In the RMF approximation all meson fields are treated as 
classical fields.  In this approximation the spatial components of the vector 
meson fields vanish. Moreover, only the third component of the $\rho$-meson 
field in isospin space survives. 
To simplify the notation, the time-like component of the 
$\omega^{0}$ field will be labelled as $\omega$. Similar notation holds 
also for the time-like and third components in Lorentz- and isospin-space, 
respectively, of the vector-isovector $\rho^{0}_{3}$ field, which will be 
denoted as $\rho$. The baryon density will be labelled as $\rho_{B}$.
For the auxiliary vector we choose $v^{\beta}=(1,\vec{0})$. 
Note that, additional rearrangement terms can appear for different choices 
of $v^{\beta}$ in the NLD formalism. However, they exactly vanish in the mean-field 
approximation to infinite nuclear matter~\cite{nld}. 

The Dirac equation, see Eq.~(\ref{Dirac}), can be solved using the following
{\it ansatz}
\begin{equation}
\Psi(s,\vec{p}\,) = u(s,\vec{p}\,)e^{-ip^{\mu}x_{\mu}} 
\left( 
\begin{array}{c}
p \\
n
\end{array}
\right)
\:, \label{plane_wave}
\end{equation}
where $s$ and $p\,^{\mu}=(E,\vec{p}\,)$ stand, respectively, for the spin and $4$-momentum of the proton ($p$) 
or neutron ($n$), $x^{\mu}$ denotes the space-time coordinate, and $u$ is the Dirac spinor  
for positive energy states. The Dirac equation (\ref{Dirac}) maintains its 
form in infinite nuclear matter with selfenergies given by
\begin{eqnarray}
\Sigma^{0}_{i}\equiv \Sigma_{vi} &=& g_{\omega}\omega \; e^{-\frac{E_{i}-m}{\Lambda}} 
\pm g_{\rho}\rho \; e^{-\frac{E_{i}-m}{\Lambda}}\,,
\label{SelfenvNM}\\
\Sigma_{si} &=& g_{\sigma}\sigma \; e^{-\frac{E_{i}-m}{\Lambda}}
\;. \label{SelfensNM}
\end{eqnarray}
The upper (lower) sign in Eq.~(\ref{SelfenvNM}) refers from now on 
always to protons, $i=p$, (neutrons, $i=n$). 
The Dirac equation is solved inserting Eqs.~(\ref{plane_wave}),
~(\ref{SelfenvNM}) and~(\ref{SelfensNM}) into Eq.~(\ref{Dirac}) 
for protons ($i=p$) or neutrons ($i=n$)
\begin{equation}
\gamma_{0}E^{*}_{i}\Psi_{i} = 
\left( 
\vec{\gamma}\cdot\vec{p}+m^{*}_{i}
\right)\Psi_{i}
\; , \label{DiracNM}
\end{equation}
with the in-medium energy and mass given by 
\begin{equation}
E^{*}_{i} := E_{i} - \left( g_{\omega}\omega e^{-\frac{E_{i}-m}{\Lambda}} 
\pm g_{\rho}\rho e^{-\frac{E_{i}-m}{\Lambda}} \right)\,,
\label{in-medium-E}
\end{equation}
and
\begin{equation}
m^{*}_{i} := m - g_{\sigma}\sigma e^{-\frac{E_{i}-m}{\Lambda}}
\;. \label{in-medium-M}
\end{equation}
These quantities are related to each other through the dispersion relation
\begin{equation}
E^{*2}_{i} - \vec{p\,}^{2} = m^{*2}_{i}
\quad ,
\label{mass-shel}
\end{equation}
which, because of the $\rho$-field, is different for protons and neutrons. The solutions of 
the Dirac equations for protons and neutrons are found in the usual way
\begin{equation}
u_i(s,\vec{p}\,) = N_i
\left(
\begin{array}{c}
\phi_{s} \\
\frac{ \vec{\sigma}\cdot\vec{p}}{E^{*}_{i}+m^{*}_{i}}\phi_{s}\\
\end{array}
\right)
\; , \label{Spinor}
\end{equation}
with the spin eigenfunctions $\phi_{s}$. The factor $N_{i}$ is defined as
$N_{i}=\sqrt{\frac{E^{*}_{i}+m^{*}_{i}}{2E^{*}_{i}}}$ and normalizes the
spinors as follows
$\bar{u}_{i}(s,\vec{p}\,)\gamma^{0}u_{i}(s,\vec{p}\,)=1$ and 
$\bar{u}_{i}(s,\vec{p}\,)u_{i}(s,\vec{p}\,)=\frac{m^{*}_{i}}{E^{*}_{i}}$.

In nuclear matter the meson field equations of motion read
\begin{equation}
m_{\sigma}^{2}\sigma = g_{\sigma}\rho_{s} 
\quad\mbox{, }\quad
m_{\rho}^{2}\rho = g_{\rho}\rho_{I}
\quad\mbox{, }\quad
m_{\omega}^{2}\omega = g_{\omega}\rho_{0}
\;, \label{mesonsNM}
\end{equation}
where the vector-isoscalar $\rho_{0}$, vector-isovector $\rho_{I}$ and 
scalar-isoscalar $\rho_{s}$ source terms take the forms~($\kappa=2$ is the
spin degeneracy factor)
\begin{eqnarray}
\rho_{s} & = & 
\sum_{i=p,n}
\langle \overline{\Psi}_{i} e^{-\frac{E_{i}-m}{\Lambda}}\Psi_{i}\rangle 
= \frac{\kappa}{(2\pi)^{3}}\; \sum_{i=p,n} \; \int 
d^{3}p \frac{m^{*}_{i}}{E^{*}_{i}} 
e^{-\frac{E_{i}-m}{\Lambda}}\Theta (p-p_{F_{i}}) = 
\rho_{sp}+\rho_{sn}
\,,
\nonumber\\
\rho_{0} & = & 
\sum_{i=p,n}
\langle \overline{\Psi}_{i} \gamma^{0} e^{-\frac{E_{i}-m}{\Lambda}}\Psi_{i}\rangle
= \frac{\kappa}{(2\pi)^{3}}\; \sum_{i=p,n} \; 
\int d^{3}p e^{-\frac{E_{i}-m}{\Lambda}}\Theta (p-p_{F_{i}}) = \rho_{0p}+\rho_{0n}
\,,
\nonumber\\
\rho_{I} & = & 
\sum_{i=p,n}
\langle \overline{\Psi}_{i}\gamma^{0}\tau_{3}e^{-\frac{E_{i}-m}{\Lambda}}\Psi_{i}\rangle = 
\rho_{0p} - \rho_{0n}
\,. \label{dens-1}
\end{eqnarray}

Contrary to the Walecka model, in the NLD approach the vector-density
$\rho_{0}$ is not a conserved nucleon density. Latter one has to be derived from 
the generalized Noether-theorem, see Eq.~(\ref{Noether-Current}). Following 
Ref.~\cite{nld} the conserved nucleon density after the resummation procedure takes the form
\begin{eqnarray}
J^{0} \equiv \rho_{B} & = & 
\sum_{i=p,n} \left[
\langle \overline{\Psi}_{i}\gamma^{0}\Psi_{i} \rangle
\right.
\label{rhoBar}\\
& - & 
\left.
\frac{g_{\sigma}}{\Lambda}
\langle \overline{\Psi}_{i}e^{-\frac{E_{i}-m}{\Lambda}} \Psi_{i} \rangle \sigma
+ \frac{g_{\omega}}{\Lambda}
\langle \overline{\Psi}_{i}\gamma^{0}e^{-\frac{E_{i}-m}{\Lambda}}\Psi_{i}\rangle\omega
+ \frac{g_{\rho}}{\Lambda}
\langle \overline{\Psi}_{i}\gamma^{0}e^{-\frac{E_{i}-m}{\Lambda}}\tau_{3}\Psi_{i}\rangle\rho
\right]
\,,
\nonumber
\end{eqnarray}
where the expectation value 
$\rho_{W}=\sum_{i=p,n}\langle \overline{\Psi}_{i}\gamma^{0}\Psi_{i} \rangle$ 
is just the usual density of the Walecka model~\cite{rhd2}. 

The equation of state (EoS) is obtained from the $00$-component of the energy-momentum 
tensor $T^{\mu\nu}$. The resummation procedure described in~\cite{nld} results in
\begin{align}
& T^{00} \equiv \epsilon =
\sum_{i=p,n}
\langle \overline{\Psi}_{i}\gamma^{0} E_{i} \Psi_{i} \rangle
\nonumber\\
& - \frac{g_{\sigma}}{\Lambda}
	\sum_{i=p,n}
	\langle \overline{\Psi}_{i}e^{-\frac{E_{i}-m}{\Lambda}} E_{i} \Psi_{i} \rangle \sigma
 +  \frac{g_{\omega}}{\Lambda}
	\sum_{i=p,n}
	\langle \overline{\Psi}_{i}\gamma^{0}e^{-\frac{E_{i}-m}{\Lambda}} E_{i} \Psi_{i} \rangle \omega
+ \frac{g_{\rho}}{\Lambda}
	\sum_{i=p,n}
	\langle \overline{\Psi}_{i}\gamma^{0}e^{-\frac{E_{i}-m}{\Lambda}} E_{i} \tau_{3} \Psi_{i} \rangle \rho
\nonumber\\
& + \frac{1}{2}
\left( 
	  m_{\sigma}^{2}\sigma^{2} - m_{\omega}^{2}\omega^{2} 
	- m_{\rho}^{2}\rho^{2}
\right)
\;, \label{eos}
\end{align}
with the additional expectation values 
\begin{eqnarray}
\rho_{s}^{E} & = & 
\sum_{i=p,n}
\langle \overline{\Psi}_{i}E_{i} e^{-\frac{E_{i}-m}{\Lambda}} \Psi_{i}\rangle 
= \frac{\kappa}{(2\pi)^{3}}\; \sum_{i=p,n} \;
\int d^{3}p \frac{m^{*}_{i}}{E^{*}_{i}} E_{i} 
            e^{-\frac{E_{i}-m}{\Lambda}}\Theta (p-p_{F_{i}}),
\nonumber\\
\rho_{0}^{E} & = & 
\sum_{i=p,n}
\langle \overline{\Psi}_{i} E_{i} \gamma^{0} e^{-\frac{E_{i}-m}{\Lambda}} \Psi_{i} \rangle
= \frac{\kappa}{(2\pi)^{3}}\; \sum_{i=p,n} \;
\int d^{3}p E_{i} e^{-\frac{E_{i}-m}{\Lambda}}\Theta (p-p_{F_{i}})\,,
\nonumber\\
\rho_{I}^{E} & = & 
\sum_{i=p,n}
\langle \overline{\Psi}_{i} E_{i} 
\gamma^{0} e^{-\frac{E_{i}-m}{\Lambda}} \tau_{3} \Psi_{i} \rangle
= \rho_{0p}^{E} - \rho_{0n}^{E}
\;. \label{dens-2}
\end{eqnarray}
The pressure $P$ is given as usual by 
\begin{equation}
P=\frac{1}{3}\left( T^{xx}+T^{yy}+T^{zz} \right). 
\end{equation}
However, since in the 
NLD approach rearrangement terms do not appear~\cite{nld}, thermodynamic consistency 
is always fulfilled. Then, the pressure can be obtained from the 
energy density by using the thermodynamic relation
\begin{equation}
P = \rho_{B}^{2}\frac{\partial\left( \epsilon/\rho_{B} \right)}{\partial\rho_{B}}
\quad .
\label{Press}
\end{equation}
Note that in Eq.~(\ref{Press}) the conserved baryon density $\rho_{B}$ from 
Eq.~(\ref{rhoBar}) appears, and not the standard Walecka density.

Two important effects appear in NLD model, which are absent in the minimal 
$\sigma\omega\rho$-models. At first, the selfenergies depend not only 
on density, but also explicitly on single-particle energy, as discussed 
in detail in Ref.~\cite{nld}. These feature is retained also in isospin-asymmetric 
nuclear matter. As an additional novel feature of the NLD model, 
not only the Lorentz-vector component of the nucleon selfenergy, 
but also its Lorentz-scalar part depends implicitly (through the dispersion
relation) on the isovector $\rho$-meson field. 
This feature arises from the isospin dependence of the single-particle 
energy, and it will generate an in-medium mass splitting between protons and 
neutrons in isospin asymmetric  matter. Note that the $\sigma$ field 
is also affected by isospin effects, but the residual impact of the dispersion 
relation is negligibly small in this case.

\section{\label{sec4}Numerical realization}

The meson-field equations depend explicitly on the baryon density, but 
only implicitly on the single-particle energy. Latter quantity is 
integrated out in the source terms. Therefore, it is sufficient to solve 
selfconsistently only the coupled set of meson-field equations 
(\ref{mesonsNM}) at a given density, 
and then determine the proton or neutron single-particle energy by 
solving numerically the dispersion relation~(\ref{mass-shel}) for a 
given proton or neutron momentum relative to the nuclear matter at rest.

For the density dependence we proceed as follows: the input parameters 
are the Fermi-momentum and the isospin-asymmetry.  The isospin-asymmetry 
parameter $\alpha$ is determined according to the standard definition 
\begin{equation}
\alpha = \frac{J^{0}_{n}-J^{0}_{p}}{J^{0}_{n}+J^{0}_{p}}
\quad . 
\label{alpha}
\end{equation}
The set of the meson field equations (\ref{mesonsNM}) is solved then 
selfconsistently until convergence is achieved. For each iteration 
step the integrals in Eq.~(\ref{dens-1}) are evaluated numerically, where 
at each momentum step the dispersion relation, see Eq.~(\ref{mass-shel}), has 
to be calculated also numerically. The whole procedure provides the meson-fields, 
with which we calculate the corresponding conserved baryon density according 
Eq.~(\ref{rhoBar}) and the isospin-asymmetry parameter according 
Eq.~(\ref{alpha}). Note again, that the conserved baryon density $\rho_{B}$ 
differs from the Walecka one $\rho^{W}$. Finally, the single-particle 
energies, which appear explicitly in the selfenergies, are evaluated 
at the corresponding proton or neutron Fermi-momentum. 

For the energy dependence at a fixed baryon density the procedure is similar as 
outlined above. We first calculate the meson-fields at a given density and 
isospin-asymmetry. Then we solve numerically the dispersion relation 
for protons (neutrons). In the first case the additional input parameter is 
the proton (neutron) momentum by keeping the momentum of the other isospin state 
at the corresponding fixed Fermi-momentum. The in-medium energy of a 
proton ($i=p$) or neutron ($i=n$) is defined then as
\begin{equation}
E_{i} = \sqrt{m^{*2}_{i}+p^{2}} + \Sigma_{vi}
\quad ,
\label{KinEn}
\end{equation}
where the proton or neutron Lorentz-vector selfenergy is given by Eq.~(\ref{SelfenvNM}). 

In principle, the NLD model contains no parameters. 
The original isoscalar $\sigma NN$ and $\omega NN$ couplings can be taken from 
any linear Walecka model, e.g., \cite{rhd1}, as it has been done in 
Ref.~\citep{nld} and in this work. 
The $\rho NN$ coupling $g_{\rho}=3$ close to its universal value
is used~\cite{Kaskulov:2011ab} which is also well determined by the one-boson-exchange 
model~\cite{obe}. 
A small (big) numerical value of $\Lambda$ 
provides a stronger (weaker) density and energy dependence of the mean-field, which 
influences the density dependence of the equation of state for densities at and above saturation 
and, at the same time, strongly affects the energy dependence of the Schr\"odinger-equivalent 
optical potential at saturation density. The simplest way followed here is to fix $\Lambda$ 
just from the empirically well-known energy dependence of the Schr\"odinger equivalent 
optical potential. The cut-off parameter $\Lambda$ is set to $770$~MeV~\cite{nld}. 
A different numerical value for the cut-off parameter is of course possible if one uses a 
different Walecka-like parametrization than the original one or if one adopts a different 
functional form for the non-linear operator than the simpler exponential one.

\section{\label{sec5}Results and discussions}

\begin{figure}[t]
\unitlength1cm
\begin{picture}(18.,8.0)
\put(0.25,0.0){\makebox{\epsfig{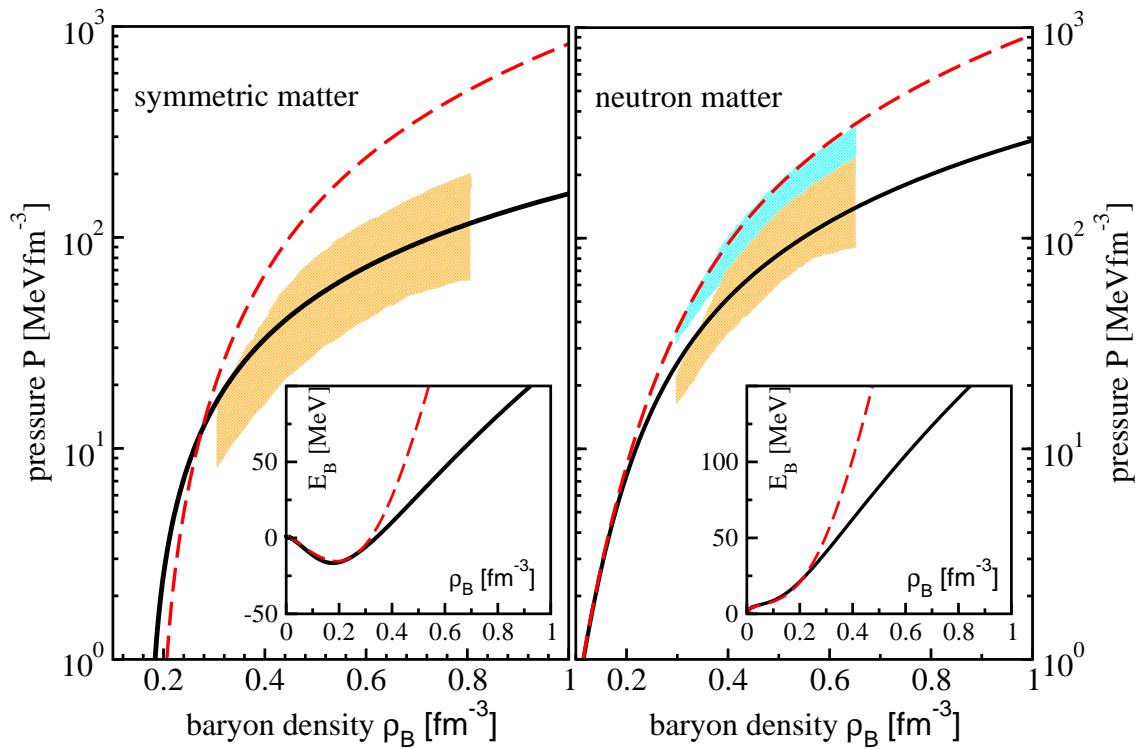}}}
\end{picture}
\caption{The nuclear EoS for symmetric nuclear matter (main panel on the left) 
and pure neutron matter (main panel on the right) in terms 
of the pressure density in the Walecka (dashed curve) and NLD (solid curve) models. 
The shaded bands refer to the experimental regions extracted 
from heavy-ion collisions~\protect\cite{ns1,dani}. The inserted panels on the 
left (right) describe the 
nuclear EoS for symmetric matter (left) and neutron matter (right) 
in terms of the binding energy per nucleon 
$E_{\rm B}=\epsilon/\rho_{B}-m$, again for both NLD and 
Walecka models. 
}
\label{fig1}
\end{figure}

Before presenting the calculations for isospin asymmetric nuclear matter 
we discuss briefly the NLD results for 
the special cases of symmetric matter ($\alpha=0$) and 
pure neutron matter ($\alpha=1$). 
In Fig.~\ref{fig1} we present the EoS for $\alpha=0$ and $\alpha=1$ 
in terms of the pressure density as function of the baryon density.
Here we compare the empirical heavy-ion data~\cite{dani}, see the shaded bands in
Fig.~\ref{fig1}, with NLD and Walecka models. The inserted panels in Fig.~\ref{fig1} 
shows the 
nuclear EoS in terms of the binding energy per nucleon. The conventional 
Walecka model (dashed curves) leads to a very stiff EoS 
for symmetric nuclear matter and also pure neutron matter 
with a high compression 
modulus at saturation, because of the well-known linear divergent behaviour 
of the repulsive $\omega$ field. On the other hand, the Lorentz-vector mean-field is 
suppressed to large extend at high baryon densities in the NLD model. 
See further discussions in Ref.~\cite{nld}. This effect, 
which is retained also for the extreme case of pure neutron matter, 
weakens considerably the stiffness of the high-density EoS, 
see the insert in Fig.~\ref{fig1}, and the empirical 
region in the pressure-density phase diagram is reproduced fairly well by 
the NLD approach (solid curve). 

However, not only the density dependence of the nuclear mean-field, but also its 
energy dependence matters for nuclear reactions. This can be demonstrated  
using the in-medium optical potential which is defined as follows
\begin{equation}
 U_{\rm opt} = \frac{E}{m} \Sigma_{v} - \Sigma_{s}
 + \frac{1}{2m} \left( \Sigma^{2}_{s} - \Sigma_{v}^{2}\right)
\:. \label{V_opt}
\end{equation}
\begin{figure}[t]
\unitlength1cm
\begin{picture}(18.,8.0)
\put(2.0,0.0){\makebox{\epsfig{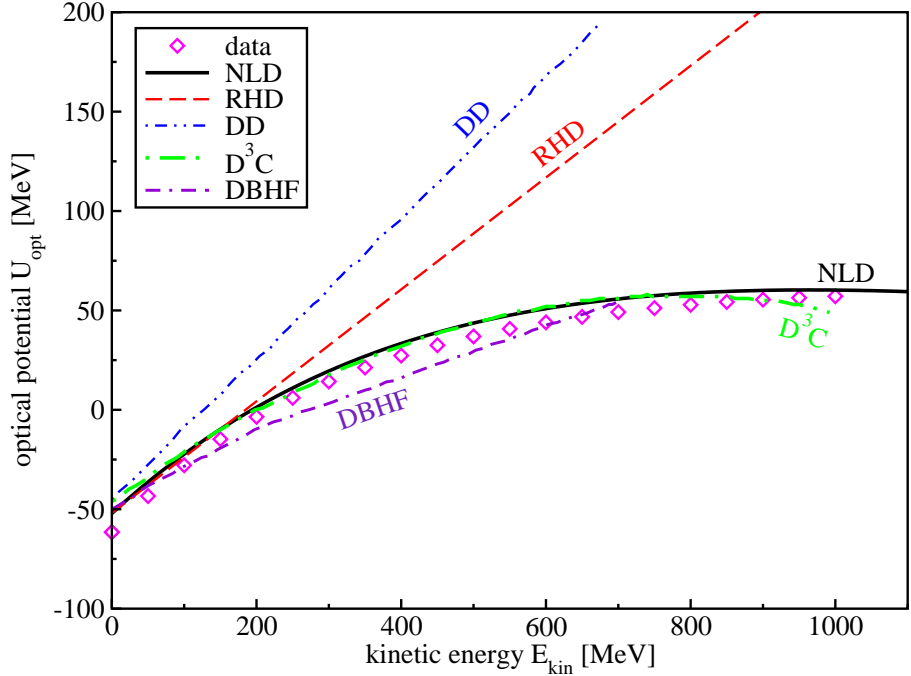}}}
\end{picture}
\caption{Kinetic energy dependence $E_{kin}=E-m$ of the Schr\"{o}dinger equivalent optical 
potential at ground state density $\rho_{sat}=0.16~fm^{-3}$ in the linear Walecka 
(RHD, dashed curve) and NLD (solid curve) models. The other theoretical 
calculations, DD (dot-dot-dashed curve), $\rm D^{3}C$ (dot-dashed curve) and 
DBHF (dot-dashed-dashed curve), are taken from Refs.~\protect\cite{typel2,dbhf2}. 
All model calculations are compared with data from 
Dirac phenomenology (open diamonds) \protect\cite{dp}. 
}
\label{fig2}
\end{figure}
As shown in Fig.~\ref{fig2}, the empirical data from
Dirac phenomenology (open diamonds) predict a 
saturation of the in-medium nucleon optical potential~\cite{dp} at intermediate 
energies. As well known, standard Walecka-type RMF models (RHD, dashed curve) are not 
able to reproduce this empirical saturation, and strongly overestimate the data at
high energies. This disagreement remains also in other RMF models 
($\rm DD$, dot-dot-dashed curve), where the meson-nucleon couplings 
depend explicitly on the baryon density. A better description can be achieved here 
by the consideration of additional energy dependent terms in the interaction 
Lagrangian ($\rm D^{3}C$, dot-dashed curve). However, with the cost 
of many additional parameters~\cite{typel2}. Microscopic DBHF models 
(DBHF, dot-dashed-dashed curve) 
reproduce the optical potential in average up to energies close to pion production 
threshold~\cite{dbhf1,dbhf2}. On the other hand, the NLD model (NLD, solid curve)
not only softens the EoS at high densities, 
but simultaneously generates the correct energy behaviour of the nucleon-nucleus optical potential 
at high energies. Note that, the same NLD approach describes the puzzling
energy dependence of the antiproton-nucleus potential, see Ref.~\cite{antinld}. 

\begin{figure}[t]
\unitlength1cm
\begin{picture}(18.,8.0)
\put(2.0,0.0){\makebox{\epsfig{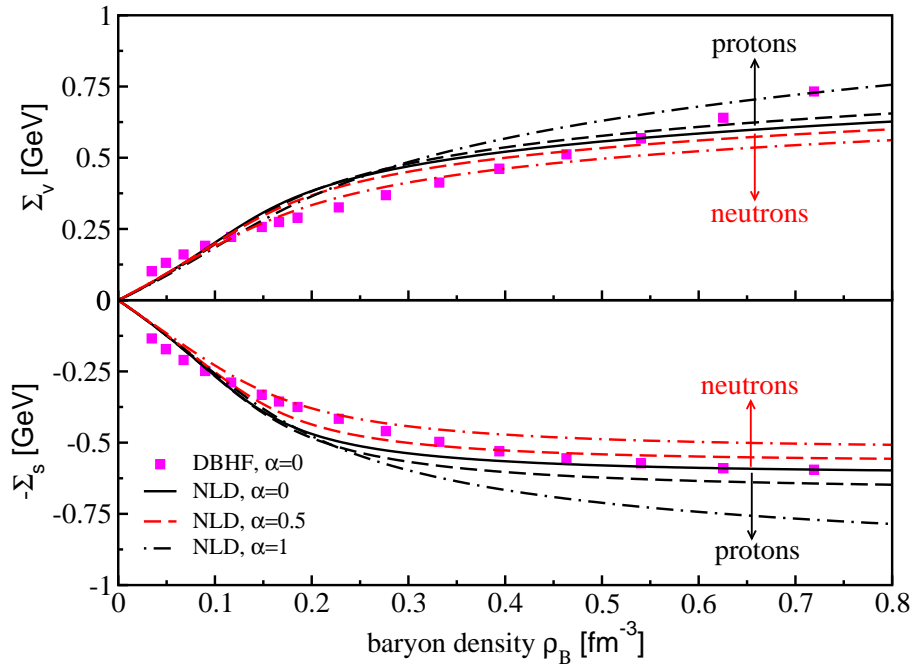}}}
\end{picture}
\caption{Density dependence of the 
vector $\Sigma_v$ (upper panel) and scalar $\Sigma_s$ (lower panel)
selfenergies. The NLD results for 
symmetric nuclear matter (solid curve) are compared with DBHF calculations~\cite{brockmann} 
(filled squares). The NLD fields for $\alpha=0.5$ (dashed curves) 
and $\alpha=1$ (dot-dashed curves) are shown too, and the direction of 
the proton and neutron splitting is marked by the arrows, as indicated.
}
\label{fig3a}
\end{figure}

In Fig.~\ref{fig3a} we present the density dependence of the nucleon selfenergies for
various asymmetry parameters $\alpha$.
In symmetric nuclear matter ($\alpha=0$), the NLD
vector $\Sigma_v$ and scalar $\Sigma_s$ selfenergies (solid curves) saturate
at high densities. As one can see, they essentially follow the density dependence of the 
microscopic DBHF calculations~\cite{brockmann} (filled squares). 
As discussed in detail in~\cite{nld}, this saturation of the 
selfenergies with increasing density, in particular, of the repulsive vector part, 
is responsible for a softening of the EoS at high compressions.
\begin{figure}[t]
\unitlength1cm
\begin{picture}(18.,8.0)
\put(2.0,0.0){\makebox{\epsfig{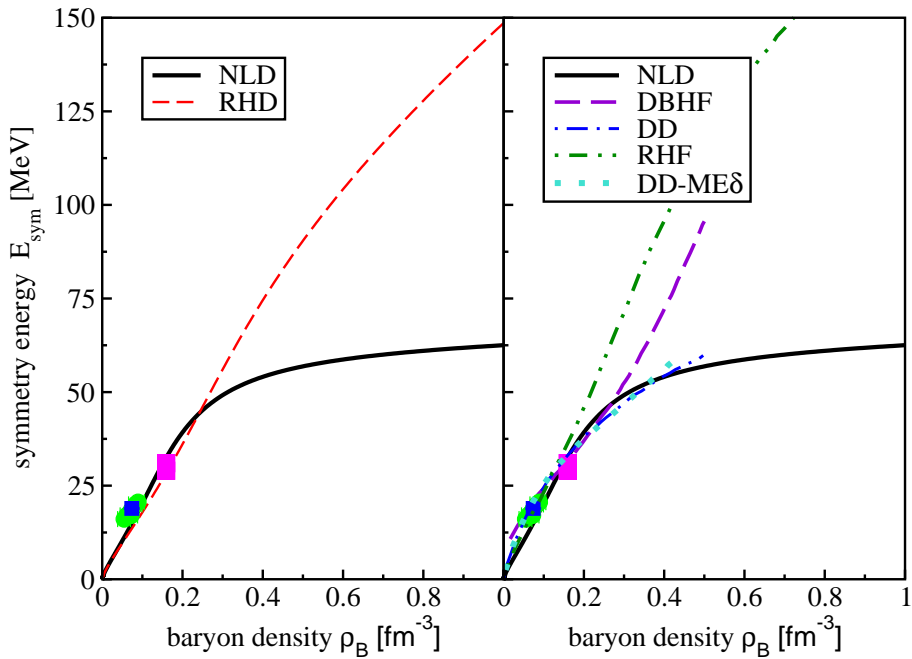}}}
\end{picture}
\caption{(Left panel) Density dependence of the symmetry energy 
$E_{\rm sym}$ in the standard linear Walecka model (dashed curve) 
and NLD approach (solid curve). (Right panel) The same as in 
the left panel, but in comparison between the NLD model with 
other theoretical approaches: DBHF (dashed curve)~\cite{dbfuchs}, 
DD (dot-dashed curve)~\cite{typel1,typel2}, RHF (dot-dot-dashed curve)~\cite{rhf2} 
and DD-ME$\delta$ (thick-doted curve)~\cite{dde}. 
The symbols refer to experimental data from Ref.~\cite{shetty1}.
}
\label{fig3}
\end{figure}
The saturation in density remains in the general case of isospin-asymmetric 
nuclear matter, see the dashed $(\alpha=0.5)$ and dot-dashed $(\alpha=1)$ 
curves in Fig.~\ref{fig3a}. As a novel feature, 
not only the vector, but also the scalar component of the 
nucleon selfenergy gets isospin dependent. The isospin dependence of the 
scalar selfenergy, observed here, arises from the $\rho$-field induced 
isospin splitting between the in-medium proton and neutron single particle
energies. This effect will be discussed in more details later on.

The saturation of the selfenergies in density affects considerably 
the density dependence of the symmetry energy. The symmetry energy is extracted from 
the empirical parabolic low and defined as the second derivative of the energy 
per nucleon versus the asymmetry parameter $\alpha$, or as the energy 
difference between pure neutron matter and symmetric nuclear matter. 
Fig.~\ref{fig3} (left panel) shows the density dependence of the symmetry 
energy within the NLD (solid curve) and the Walecka (dashed curve) 
models. The symmetry energy 
$E_{sym}=T_{sym}+V_{sym}$ in the standard RMF grows almost linearly with 
baryon density. Here the symmetry potential $V_{sym}$ is  
proportional to the $\rho$ field, which increases linearly with density,  
$\rho_{B}$. The moderate saturation at very high densities arises from the 
suppression of the kinetic part $T_{sym} \sim \frac{\rho^{2/3}}{E_{F}^{*}}$ 
(with $E_{F}^{*}$ being the in-medium Fermi-energy). 
The NLD model gives a different behaviour at high baryon densities with 
respect to the conventional RMF. A considerable softening 
of the high density EoS is obtained, but now for the isovector sector. 

Available data for the symmetry energy exist around saturation 
density only~\cite{shetty1}. The theoretical models describe the
experimental data fairly well, 
however, they give different predictions for the high-density symmetry energy, 
see for reviews Refs.~\cite{ditoro,baoan}. In non-relativistic models 
the adjustment of the model parameters to the empirical saturation point does 
not necessarily constraint the stiffness of the symmetry energy. In fact, 
a soft and a stiff symmetry energy can be obtained at high 
densities~\cite{baoan}. In the relativistic Walecka-type models, the high density 
dependence of the symmetry energy is always stiff~\cite{kubis,liu}. 
Similar results are obtained in the 
DBHF approach and within density-dependent relativistic Hartree-Fock (RHF) 
approaches~\cite{rhf1,rhf2}, 
see the dashed and dot-dot-dashed curves in the  right panel of Fig.\ref{fig3}, 
where the symmetry energy becomes stiffer with increasing density~\cite{dbfuchs}. 
It is interesting to note that the RHF models contains also an energy dependence, 
apart the density dependent coupling functions. 
On the other hand, the DD models~\cite{typel1,typel2,dde}, see the dot-dashed (DD) 
and dot (DD-ME$\delta$) curves in the right panel of  Fig.\ref{fig3}, predict 
a rather soft symmetry energy, which is similar to the NLD results (solid curve).

Direct experimental access to the high-density region of the symmetry energy is 
still incompletely known. A plausible way to constraint the stiffness of 
the symmetry energy at supra-normal densities consists in studies of compact 
neutron stars~\cite{ns1}. In fact, the symmetry energy must saturate or, at least, 
show a rather soft behaviour at high baryon densities in order to respect the 
direct URCA-limit~\cite{ns1}. This constraint is satisfied in the NLD model. 

\begin{figure}[t]
\unitlength1cm
\begin{picture}(18.,8.0)
\put(2.0,0.0){\makebox{\epsfig{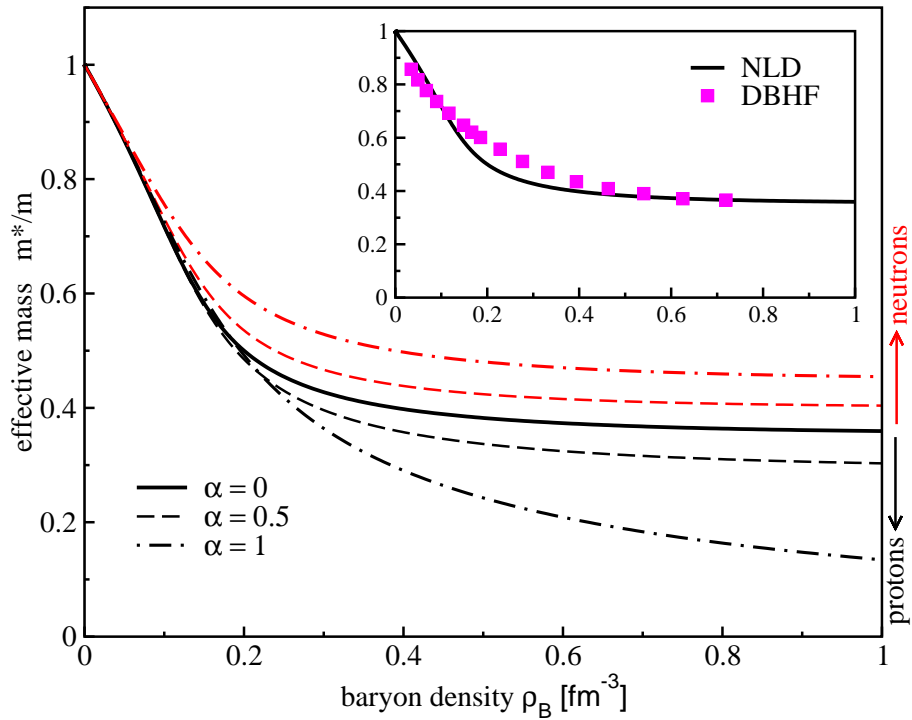}}}
\end{picture}
\caption{Density dependence of the Dirac mass splitting between protons and 
neutrons within the NLD model. The asymmetry parameters are $\alpha=0$ 
(solid curve), $\alpha=0.5$ (dashed curves) and $\alpha=1$ (dot-dashed curve). 
The curves above (below) the solid curve (symmetric matter) correspond to 
the neutron (proton) effective masses, as indicated.  The insert describes 
the Dirac mass of the nucleon in symmetric nuclear matter in the 
NLD (solid) and DBHF~\cite{brockmann} (filled squares) calculations.
}
\label{fig4}
\end{figure}

The density dependence of the nucleon effective or Dirac  mass 
in the NLD model~\cite{nld} is in
remarkable agreement with the DBHF calculations. The 
insert in Fig.~\ref{fig4} describes the Dirac mass of the nucleon in symmetric
nuclear matter. Here the solid curve corresponds to the NLD results and the square
symbols describe the DBHF calculations of Ref.~\cite{brockmann}. 

Interestingly, in the NLD model  
an effective mass splitting between protons and neutrons is 
generated in asymmetric matter by taking into account at the Hartree-level 
the $\rho$-field only. This is demonstrated 
in Fig.~\ref{fig4} for two asymmetry parameters $\alpha=0.5$~(dashed curves)
and $\alpha=1$~(dot-dashed curves). 
In particular, a mass splitting $m^{*}_{n} > m^{*}_{p}$ is obtained for
neutron-rich matter, 
which becomes pronounced with increasing density and isospin asymmetry parameter. 
The in-medium mass splitting in the NLD model is due to the isospin dependence of the single
particle energy which appears in the selfenergies. Note that, any explicit energy
dependence of the scalar selfenergy will result in a $\rho$-induced mass
splitting of nucleons in isospin asymmetric nuclear matter. 

As discussed in the introduction, an effective mass 
splitting between protons and neutrons do not occur in the Walecka-type 
$\sigma\omega\rho$-models solved at the Hartree-level. 
The introduction of the scalar-isovector $\delta$-meson in the nuclear mean-field 
is necessary for this purpose~\cite{kubis,liu}. However, one gets more parameters 
to fit the bulk asymmetry parameter at saturation density and, 
in particular, one needs to re-adjust again the $\rho NN$ coupling. This procedure 
can be ambiguous, since the $\delta NN$ coupling is 
not fixed neither by nuclear structure~\cite{typel1} nor by the one-boson-exchange 
models~\cite{brockmann}. On the other hand, as it has been investigated intensively 
in the past, the $\delta$-meson implies interesting effects if one goes beyond 
the saturation density of ground state nuclear matter and study its influence in 
compressed matter created in heavy-ion collisions~\cite{mesrat0,mesrat1}. Unfortunately, 
the available experimental data do not allow still decisive conclusions~\cite{lopez}. 
The $\sigma\omega\rho\delta$-models predict a mass splitting  
$m^{*}_{n} < m^{*}_{p}$ for neutron-rich
matter~\cite{ditoro,kubis,liu,mesrat0}. However,
just from the energy dependence of the scalar selfenergy
the NLD model predicts an opposite splitting $m^{*}_{n} > m^{*}_{p}$ for 
neutron-rich matter. This trend is unique and cannot be reversed in the NLD model. 
We remind here, that RMF models treated on the Hartree-Fock 
level (RHF)~\cite{rhf1,rhf2}, in which the selfenergies depend also explicitly 
on the single-particle energy due to the Fock-terms, predict for the 
non-relativistic effective mass a mass splitting 
$m^{*}_{n} > m^{*}_{p}$ up to densities $\rho_{B}<0.8\rho_{sat}$, 
but this trend is reversed for densities $\rho_{B} > 0.8\rho_{sat}$~\cite{rhf1}. 
Therefore, the property $m^{*}_{n} < m^{*}_{p}$ is not a genuine feature 
of the RMF models, as has been believed so far. 

\begin{figure}[t]
\unitlength1cm
\begin{picture}(18.,8.0)
\put(2.0,0.0){\makebox{\epsfig{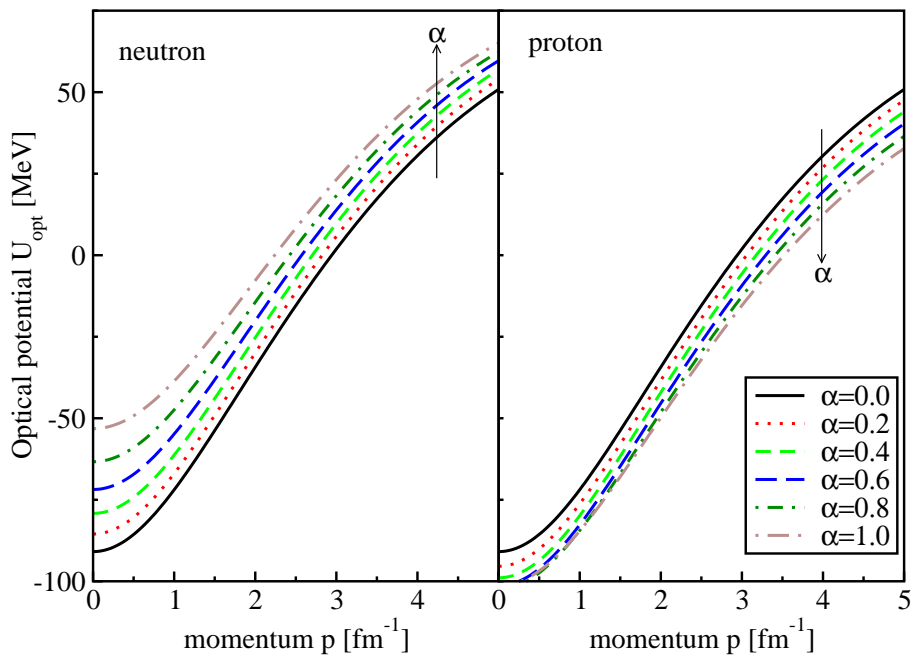}}}
\end{picture}
\caption{Momentum dependence of the Schr\"odinger equivalent neutron (left panel) 
and proton (right panel) optical potentials at various asymmetries, as indicated. 
Calculations in the NLD model are shown only. The standard RHD model does not contain 
a momentum dependence.
}
\label{fig5}
\end{figure}

\begin{figure}[t]
\unitlength1cm
\begin{picture}(18.,8.0)
\put(2.0,0.0){\makebox{\epsfig{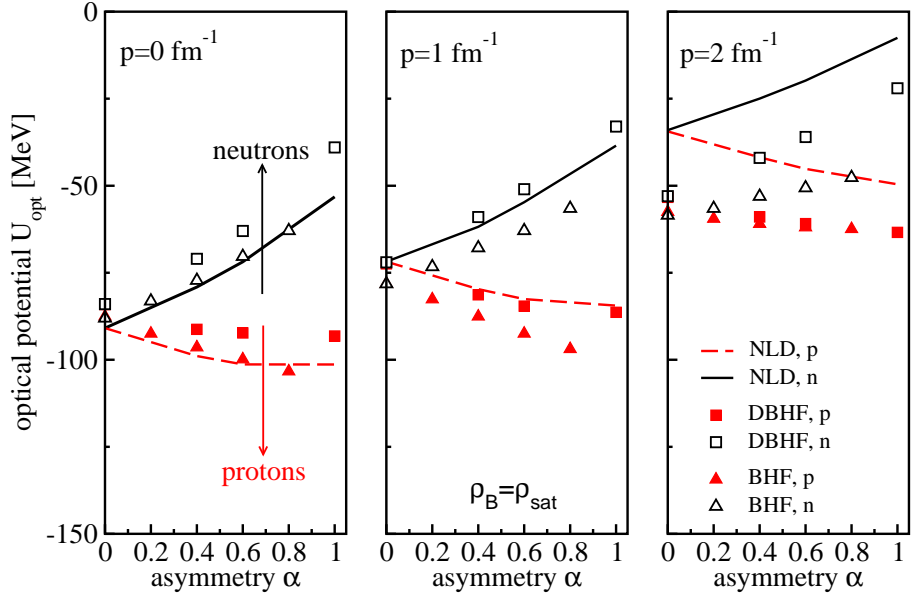}}}
\end{picture}
\caption{Asymmetry parameter $\alpha$ dependence of the optical potentials at fixed 
saturation density and at three fixed momenta ($p=0,1,2~fm^{-1}$ as indicated) 
in the NLD (solid and dashed curves for neutrons and protons), in the 
DBHF~\protect\cite{dbfuchs} (filled and open squares for protons and neutrons) and in the 
BHF~\cite{bhf1} (filled and open triangles for protons and neutrons) models. The 
arrows in the left panel indicate the direction of the proton and 
neutron splitting in these three figures.
}
\label{fig6}
\end{figure}
The energy dependence of the isovector 
nuclear potential is described by the Schr\"odinger-equivalent 
optical potential, see Eq.~(\ref{V_opt}), 
but now separated between protons and neutrons. 
Fig.~\ref{fig5} shows these potentials as function of the corresponding 
particle momentum and different asymmetry parameters $\alpha$.   
The observed isospin dependence of the 
proton (right panel) and neutron (left panel) optical potentials 
results from cancellation effects in Eq.~(\ref{V_opt}) between different
isospin dependent components of the selfenergies. 
It is remarkable that the microscopic (D)BHF calculations~\cite{dbfuchs,bhf1,bhf2} 
predict very similar
behaviour as the NLD model. This is shown in Fig.~\ref{fig6}, where 
the proton and neutron optical potentials are shown as a function of the 
asymmetry parameter $\alpha$ at fixed density $\rho=\rho_{sat}$ and three different momenta, 
as indicated in Fig.~\ref{fig6}. Indeed, at low momenta up to the Fermi momentum 
(left and middle panels) the isospin splitting between the proton and 
neutron optical potentials in the NLD model (solid and dashed curves) follows 
closely the microscopic calculations (filled and open squares for DBHF, filled and open 
triangles for BHF). At higher momenta (right panel) 
the NLD isospin splitting is similar to the (D)BHF calculations, up to an absolute 
value. This is related to a stronger energy dependence in the NLD model 
relative to these microscopic approaches. Recall, that this strong energy 
dependence of the NLD potential is essential for the description of the 
Schr\"odinger-equivalent optical potential shown in Fig.~\ref{fig2}. 

\begin{figure}[t]
\unitlength1cm
\begin{picture}(18.,8.0)
\put(2.0,0.0){\makebox{\epsfig{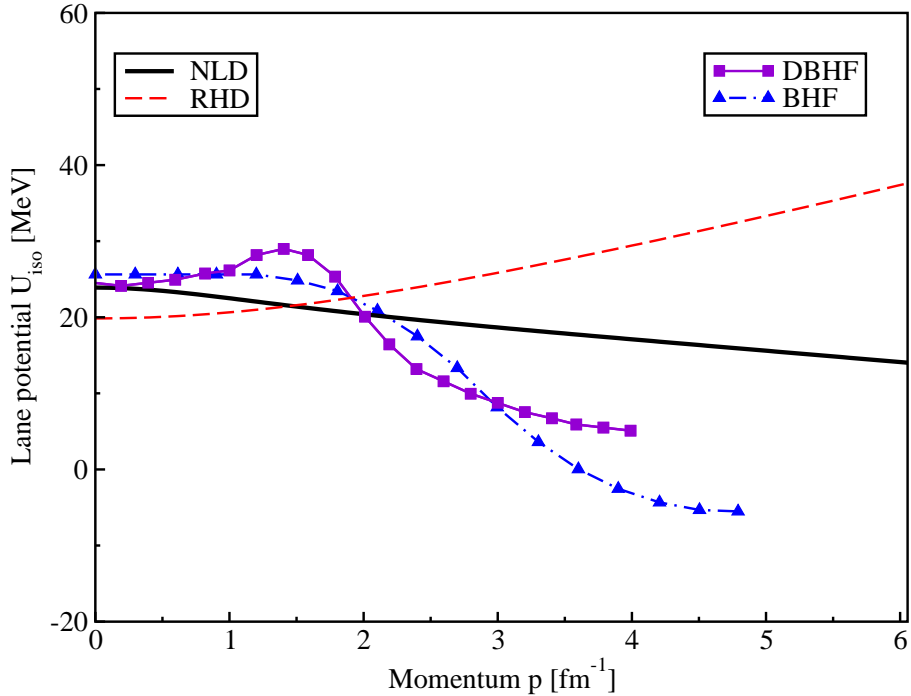}}}
\end{picture}
\caption{Energy dependence of the Lane potential at saturation density 
and an isospin asymmetry parameter $\alpha=0.4$. The NLD model (solid curve) 
is compared with the Walecka model (RHD, dashed curve) and with the 
DBHF~\cite{dbfuchs} (solid curve with filled squares) and BHF~\cite{bhf1,bhf2} 
(dot-dashed curve with filled triangles) models.
}
\label{fig7}
\end{figure}
The energy dependence of the isovector optical potential is conventionally 
described by the Lane-type potential. It is defined as follows~\cite{ditoro,baoan}
\begin{equation}
U_{\rm iso} = \frac{U_{n}-U_{p}}{2|\alpha|}
\, ,
\label{Lane}
\end{equation}
where $U_{p,n}$ denotes the proton/neutron Schr\"odinger-equivalent optical potentials. 
Fig.~\ref{fig7} shows the momentum dependence of the Lane-potential in the NLD
approach (solid curve). Also the results of 
linear Walecka model denoted as RHD (dashed curve) and different microscopic 
approaches are shown. The linear Walecka model predicts an almost linear behaviour in energy, 
as in the isoscalar case, and it diverges for $p\to\infty$. The Lane potential in 
the NLD model shows an opposite behaviour with respect to RHD. It 
decreases with increasing particle momentum. In contrast to the symmetric 
proton- and antiproton-nucleus optical potentials~\cite{nld,antinld}, the empirical situation in
the isovector case is still 
not well known. Different empirical studies predict an opposite 
behaviour of $U_{\rm iso}$ versus momentum. In fact, the analysis of Madland 
{\it et al.}~\cite{isodp2,isodp2b} 
predicts an increase of the isovector potential with increasing momentum, while in the 
analysis of Lane {\it et al.}~\cite{isodp1,isodp1b,isodp1c} a decrease of $U_{\rm iso}$ 
with rising momentum has been obtained. 
Nevertheless, as shown in 
Fig.~\ref{fig7}, the NLD calculations agree  
with the calculations from the microscopic Brueckner theory which also show
the decrease of the Lane potential as a function of the momentum.

\section{\label{sec7}Summary}

In summary, the NLD formalism which incorporates on a mean-field level 
the energy dependence of the nucleon selfenergies has been applied 
to isospin-asymmetric nuclear matter. The model describes the bulk properties of 
symmetric matter and compares well with the results from microscopic Brueckner
calculations. Due to the explicit energy dependence of the nuclear mean-field,
the NLD model reproduces very well the empirical proton and antiproton optical 
potentials~\cite{nld,antinld}. 
Same energy dependence is responsible for the softening of the EoS at high
densities. 
In the present model  the isospin effects are generated by using the
conventional $\rho$-meson exchange. Such a minimal extension
of the NLD formalism describes the symmetry energy around saturation
and predict its softening (saturation) at high baryon densities. 
This feature is an important element in
the description of dense nuclear systems created in heavy ion
collisions and/or in the interior of neutron stars.   
However, such a soft symmetry energy is at variance with standard RHD models, where 
it  rises linearly with increasing density. 

As a novel feature we find that the energy dependent NLD interactions generate a mass
splitting between protons and neutrons already within the conventional $\rho$-exchange 
scheme.
In particular, a Dirac mass splitting $m^{*}_{n} > m^{*}_{p}$ is obtained 
for neutron-rich matter. However, we  point out that the $\rho$-induced
nucleon mass splitting is a rather general feature of energy dependent
interactions and should show up in any model which incorporates them. 
 
We further discussed the energy dependence of the isovector optical
potentials. We found that the NLD model leads to a decreasing Lane-type potential
with increasing particle momentum. This disagrees with the conventional RMF
models, but is consistent with the results from microscopic Brueckner
calculations. Also the comparison of the isovector splitting in the proton/neutron 
optical potentials agrees well with the results from DBHF and BHF
microscopic approaches. These suggest that the NLD $\sigma\omega\rho$ mean-field
model which accommodates the energy dependence of nucleon
selfenergies, 
is in qualitative agreement with microscopic (Dirac) Brueckner calculations. 
Since the NLD model describes also asymmetric matter features 
fairly well and agrees with microscopic DBHF models, it could be 
interesting to apply the NLD approach to finite nuclei. However, 
due to the energy dependence of the fields, 
the main advantage of the NLD approach would be its application 
to the description of heavy-ion reactions within the transport theory.

\section*{Acknowledgments}
This work was supported by HIC for FAIR, DFG through TR16
and by BMBF.

\section*{References}





\end{document}